\documentclass[12pt,english]{article}
\usepackage[T1]{fontenc}
\usepackage[latin1]{inputenc}
\usepackage{array}
\usepackage{amsmath}
\usepackage{amssymb}
\usepackage[numbers]{natbib}

\makeatletter

\providecommand{\tabularnewline}{\\}

\usepackage{mathrsfs}

\newcommand{\aplt}{\ {\raise-.5ex\hbox{$\buildrel<\over\sim$}}\ }
\newcommand{\apgt}{\ {\raise-.5ex\hbox{$\buildrel>\over\sim$}}\ }
\usepackage[left=2cm,top=2.1cm,right=2.1cm,bottom=1cm,nohead]{geometry}

\usepackage{setspace}

\usepackage{babel}
\makeatother

\begin{document}

\section*{\textmd{\normalsize COLO-HEP-555}}

\begin{center}
\textbf{\Large A Comparison of Supersymmetry Breaking and Mediation
Mechanisms}
\par\end{center}{\Large \par}

\begin{center}
\vspace{0.3cm}
\par\end{center}

\begin{center}
{\large S.P. de Alwis{*}} 
\par\end{center}

\begin{center}
Physics Department, University of Colorado, \\
 Boulder, CO 80309 USA 
\par\end{center}

\begin{center}
\vspace{0.3cm}
\par\end{center}

\begin{center}
\textbf{Abstract}
\par\end{center}

We give a unified treatment of different models of supersymmetry breaking
and mediation from a four dimensional effective field theory standpoint.
In particular a comparison between GMSB and various gravity mediated
versions of SUSY breaking shows that, once the former is embedded
within a SUGRA framework, there is no particular advantage to that
mechanism from the point of view of FCNC suppression. We point out
the difficulties of all these scenarios - in particular the cosmological
modulus problem. We end with a discussion of possible string theory
realizations.

\begin{center}
\vspace{0.3cm}
\par\end{center}

\vfill

{*} email: dealwiss@colorado.edu

\eject

\section{Introduction }

In this note we will give a unified treatment of the various alternatives
that have been proposed in the literature for supersymmetry breaking
and its mediation to the visible sector. The latter will be taken
to be the minimally supersymmetric standard model (MSSM) though the
arguments can be trivially generalized to extensions of the MSSM. 

The first and oldest is mSUGRA (for a recent review see \citep{Arnowitt:2009qt}).
This theory uses the fact that a general supergravity (SUGRA) can
have chiral scalar fields which are neutral under the standard model
(SM) gauge group. Supersymmetry breaking happens when at the minimum
of the potential for these moduli, some of them will acquire SUSY
breaking values (i.e. their F-terms get non-zero values). This is
communicated to the visible sector by the moduli, which couple with
gravitational strength to the visible (MSSM) sector. Essentially the
point is that in general when the MSSM is embedded in SUGRA, the Yukawa
couplings as well as the gauge couplings will be functions of the
moduli. In the low energy theory these will essentially act as spurion
fields that generate a set of soft supersymmetry breaking terms at
some high scale - typically taken to be the GUT scale. However a generic
SUGRA will not yield a set of universal SUSY breaking parameters,
so that experimental constraints on flavor changing neutral currents
(FCNC) will rule out many such models. mSUGRA postulates that a sub-class
of models will generate such a set. One of the aims of this work is
to identify this sub-class and see whether it can be justified in
terms of string theory. The scale of the soft terms is set by the
value of the gravitino mass $m_{3/2}$. Thus if SUSY is to solve the
gauge hierarchy problem this mass should be of the order of a few
hundred $GeV$. This gives a natural explanation of the so-called
$\mu$-problem but this value of the gravitino mass is known to be
cosmologically problematic. The associated light modulus (the scalar
partner of the Goldstino) will also cause cosmological problems.

A variant of mSUGRA is a class of models in which the classical supersymmetry
breaking parameters generated in mSUGRA are suppressed, so that the
soft terms are essentially generated by quantum anomaly effects. These
are often called sequestered models. In the original version of this
scenario (called anomaly mediated supersymmetry breaking - AMSB) \citep{Randall:1998uk,Giudice:1998xp},
it was argued that both gaugino masses and slepton and squark masses
were generated by these effects. However as was pointed out in \citep{deAlwis:2008aq}
(based on the work of \citep{Kaplunovsky:1994fg}) the Weyl anomaly,
while generating gaugino masses, cannot directly affect the soft scalar
masses%
\footnote{The Weyl anomaly only affects the gauge kinetic terms (at the two
derivative level) and hence only gives a correction to the gauge coupling
superfield which then leads to Weyl anomaly generated contributions
to the gaugino mass. The latter is given by the expression for the
gaugino mass given in equation (G.2) of \citet{Wess:1992cp}, once
the Kaplunovsky Louis formula \citep{Kaplunovsky:1994fg} for the
gauge coupling function is used. The argument in the literature for
a contribution to the soft mass and the $A$-term depends on inserting
factors of the Weyl compensator into the wave function renormalization.
This has no justification whatsoever. A physical result cannot depend
on the particular formalism of SUGRA that is used and should be derivable,
for instance, in the formalism used in Wess and Bagger \citep{Wess:1992cp}
(where this particular auxiliary superfield is set equal to unity).
This is possible for the correction to the gauge coupling function
but there is no analog of this for the wave function renormalization.
For details see \citep{deAlwis:2008aq}. %
}. These can however be generated by the non-zero gaugino masses through
renormalization group (RG) evolution down from the high scale to the
weak scale - a mechanism which is usually called gaugino mediated
supersymmetry breaking (inoMSB) \citep{Chacko:1999mi,Kaplan:1999ac}.
A class of string theoretic models where this combined mechanism is
operative was discussed in \citep{deAlwis:2009fn}. The detailed phenomenology
has been worked out in \citep{Baer:2010uy} where this mechanism was
called inoAMSB. Since in this class of theories the classical soft
terms (as well as high scale quantum corrections) are suppressed and
the anomaly and RG running effects are flavor diagonal, there are
no FCNC problems. The salient feature of this class of theories is
that, since the soft parameters are typically suppressed by a loop
factor compared to $m_{3/2}$, the latter has to be taken at a scale
around $100TeV$. This avoids the cosmological gravitino problem of
mSUGRA. Nevertheless there is a cosmological modulus problem (unless
we increase the fine-tuning in the little hierarchy) as we discuss
later, and we also lose the natural solution to the $\mu$ problem.

The third class of theories is called gauge mediated supersymmetry
breaking (GMSB) (for a review see \citep{Giudice:1998bp}). In almost
all of the discussions of GMSB, the question of embedding within supergravity
is not discussed. The essential motivation for this class of theories
is the need to find a natural solution to the FCNC problem. Thus it
is postulated that supersymmetry breaking (which takes place in some
hidden sector) is transmitted to the visible sector by gauge interactions.
These theories must of course still be embedded in a supergravity
and thus in order to avoid soft parameters generated by mSUGRA or
its variants, it has to be assumed that the gravitino mass is well
below the weak scale. Typically it is taken to be below the $KeV$
scale in order to avoid cosmological problems. However now there is
no natural solution to the $\mu$ problem and furthermore there is
a $B\mu$ problem as well.

In the next section we will present the basic formulae which give
us a general framework for discussing all of these theories from a
unified perspective. In the subsequent sections we will take up these
three classes and present their theoretical underpinnings and discuss
their possible string theory origin.

\section{Generalities}

We set $M_{P}=(8\pi G_{N})^{-1}=2.4\times10^{18}GeV=1$. Let the gauge
neutral moduli (some of which will acquire non zero vacuum values
to break SUSY) be denoted by $\Phi=\{\Phi^{A}\}$. We expand the superpotential
and the Kaehler potential in powers of the chiral superfields $C^{\alpha}$
which represent the visible sector (MSSM/GUT) fields. The coefficients
of this expansion will be functions of the moduli $\Phi$. So we write
respectively, the superpotential, Kaehler potential and gauge kinetic
function for the theory under discussion as,

\begin{eqnarray}
W & = & \hat{W}(\Phi)+\frac{1}{2}\tilde{\mu}_{\alpha\beta}(\Phi)C^{\alpha}C^{\beta}+\frac{1}{6}\tilde{Y}_{\alpha\beta\gamma}(\Phi)C^{\alpha}C^{\beta}C^{\gamma}+\ldots,\label{eq:W}\\
K & = & \hat{K}(\Phi,\bar{\Phi})+\tilde{K}_{\alpha\bar{\beta}}(\Phi,\bar{\Phi})C^{\alpha}C^{\bar{\beta}}+[Z_{\alpha\beta}(\Phi,\bar{\Phi})C^{\alpha}C^{\beta}+h.c.]+\ldots\label{eq:K}\\
f_{a} & = & f_{a}(\Phi).\label{eq:f_a}\end{eqnarray}
The effective theory valid at some high scale well below the Planck/string
scale, should be such that the potential for the moduli has at least
one supersymmetry breaking minimum with nearly zero cosmological constant.
This of course requires fine-tuning of some parameters, or an appropriate
choice of internal fluxes in string theory constructions. Then the
softly broken globally supersymmetric low energy theory is described
in terms of an effective superpotential \begin{equation}
W^{(eff)}(C)=\frac{1}{2}\mu_{\alpha\beta}C^{\alpha}C^{\beta}+\frac{1}{6}Y_{\alpha\beta\gamma}C^{\alpha}C^{\beta}C^{\gamma}+\ldots,\label{eq:Weff}\end{equation}
and the SUSY breaking terms (with $c$ denoting the lowest component
of $C$)\begin{equation}
\Delta V^{(eff)}(c)=m_{\alpha\bar{\beta}}^{2}c^{\alpha}c^{\bar{\beta}}+(\frac{1}{2}(B\mu)_{\alpha\beta}c^{\alpha}c^{\beta}+\frac{1}{6}A_{\alpha\beta\gamma}c^{\alpha}c^{\beta}c^{\gamma}+h.c.).\label{eq:DeltaV}\end{equation}
The effective SUSY and SUSY breaking parameters are then given by
\citep{Kaplunovsky:1993rd} (assuming the CC is tuned to zero),

\begin{eqnarray}
\mu_{\alpha\beta} & = & e^{\hat{K}/2}\tilde{\mu}_{\alpha\beta}+m_{3/2}Z_{\alpha\beta}-\bar{F}^{\bar{A}}\partial_{\bar{A}}Z_{\alpha\beta},\label{eq:mu}\\
Y_{\alpha\beta\gamma} & = & e^{\hat{K}/2}\tilde{Y}_{\alpha\beta\gamma}\label{eq:Yukawa}\\
B\mu_{\alpha\beta} & = & F^{A}D_{A}\mu_{\alpha\beta}-m_{3/2}\mu_{\alpha\beta},\label{eq:Bmu}\\
M_{i} & = & \frac{F^{A}\partial_{A}H_{i}}{2H_{a}},\label{eq:Mgaugino}\\
m_{\alpha\bar{\beta}}^{2} & = & m_{3/2}^{2}\tilde{K}_{\alpha\bar{\beta}}-F^{A}F^{\bar{B}}R_{A\bar{B}\alpha\bar{\beta}},\label{eq:msoft}\\
A_{\alpha\beta\gamma} & = & F^{A}D_{A}Y_{\alpha\beta\gamma},\label{eq:Aterm}\end{eqnarray}
with $D_{A}\equiv\nabla_{A}+K_{A}/2$ where $\nabla_{A}$ is the usual
covariant derivative. In the above all the moduli are to be fixed
at the SUSY breaking minimum. It should be emphasized that the Kaehler
potential that is to be used in these formulae should be the effective
Kaehler potential at the high scale - i.e. it should include all quantum
corrections at that scale. On the other hand the Weyl anomaly will
change the gauge coupling function from the classical function $f(\Phi)$
to the effective gauge coupling function $H(\Phi)$. The correct formula
for this replacement is \citep{Kaplunovsky:1994fg}\begin{equation}
f_{i}\rightarrow H_{i}=f_{i}-\frac{3c_{i}}{8\pi^{2}}\tau-\sum_{r}\frac{T_{i}(r)}{4\pi^{2}}\tau_{r}-\frac{T(G_{i})}{4\pi^{2}}\tau_{i},\label{eq:Hphys}\end{equation}
where the $\tau's$ are various chiral rotations which are fixed by
the following expressions:\begin{eqnarray}
\tau+\bar{\tau} & = & \frac{1}{3}K|_{{\rm harm}},\label{eq:phi}\\
\tau_{r}+\bar{\tau}_{r} & = & \ln\det\tilde{K}_{\alpha\bar{\beta}}^{(r)},\label{eq:tauZ}\\
\exp[-(\tau_{i}+\bar{\tau}_{i})]|_{{\rm harm}} & = & \frac{1}{2}(H_{i}+\bar{H}_{i}).\label{eq:tauV}\end{eqnarray}
In the above $|_{harm}$ is an instruction to keep only the chiral
plus anti-chiral components of the relevant expressions.

The entire phenomenological content at some high scale (GUT scale/string
scale or messenger scale) of all theories of SUSY breaking and transmission
are contained in the formulae \eqref{eq:mu} to \eqref{eq:Hphys}.
In the following we will elaborate on this statement.

\section{SUGRA assumptions}

Here we will discuss the assumptions at the supergravity level that
lead to the various mechanisms of SUSY breaking and mediation. In
the next section we will examine to what extent these assumptions
can be justified from string theory.

\subsection{mSUGRA}

mSUGRA is phenomenologically defined by a set of input parameters
at some high scale - typically chosen to be the GUT scale. Thus one
chooses a universal value $m_{0}$ for the soft masses, another universal
parameter $A_{0}$ for the scalar Yukawa couplings (i.e. $A_{\alpha\beta\gamma}=A_{0}Y_{\alpha\beta\gamma}$),
and $ $ a universal gaugino mass $M$. The $\mu$ parameter, as we
observed earlier, comes out to be of the right order when we choose
$m_{3/2}\sim m$ at the weak scale, and in mSUGRA phenomenology its
exact magnitude is fixed by demanding the right $Z$ mass, leaving
the sign of $\mu$ as a parameter. $B\mu$ however is traded for $\tan\beta$,
the ratio of the two Higgs vev's. 

From a SUGRA point of view, a sufficient condition for the mSUGRA
universality choice for the scalar masses is obtained, by simply demanding
that the Kaehler metric on moduli space in the visible sector directions
is conformal to a flat (moduli independent) metric. i.e.\begin{equation}
\tilde{K}_{\alpha\bar{\beta}}=g(\Phi,\bar{\Phi})k_{\alpha\bar{\beta}},\label{eq:VmetricmSUGRA}\end{equation}
where $k_{\alpha\bar{\beta}}$ is a constant matrix. In this case
$R_{A\bar{B}\alpha\bar{\beta}}=\partial_{A}\partial_{\bar{B}}\ln g(\Phi,\bar{\Phi})K_{\alpha\bar{\beta}}$
so that from \eqref{eq:msoft} \begin{equation}
m_{\alpha\bar{\beta}}^{2}=(m_{3/2}^{2}-F^{A}\bar{F}^{\bar{B}}\partial_{A}\partial_{\bar{B}}\ln g(\Phi,\bar{\Phi}))K_{\alpha\bar{\beta}}.\label{eq:m0mSUGRA}\end{equation}
To get scalar Yukawa couplings proportional to the original ones,
a sufficient additional assumption is that the original Yukawa couplings
$\tilde{Y}$ are independent of the SUSY breaking moduli %
\footnote{Note that these conditions are considerably weaker than what is usually
assumed as being a set of sufficient conditions for mSUGRA - see for
example \citep{Dine:2009gy}.%
}. In this case (since from \eqref{eq:VmetricmSUGRA} $\Gamma_{A\beta}^{\alpha}=\partial_{A}\ln g\delta_{\beta}^{\alpha}$)
we have from \eqref{eq:Aterm}\begin{equation}
A_{\alpha\beta\gamma}=F^{A}(\frac{1}{2}K_{A}+\partial_{A}\ln g(\Phi,\bar{\Phi}))Y_{\alpha\beta\gamma}\equiv A_{0}Y_{\alpha\beta\gamma}.\label{eq:A0mSUGRA}\end{equation}
 Finally to get universal gaugino masses (as is usually assumed in
mSUGRA) one needs to assume that the gauge coupling functions of the
three factors of the standard model gauge group have the same dependence
on the (SUSY breaking) moduli. However although this assumption can
in fact be realized in some string theoretic constructions, it is
not crucial since non-universal gaugino masses at the high scale do
not violate any phenomenological constraint. But leaving that aside,
the simple assumption \eqref{eq:VmetricmSUGRA} and the assumption
of $\Phi$ independence of the Yukawa couplings (for $\Phi$'s which
break SUSY at the $m_{3/2}$ scale, give a viable phenomenology and
a testable set of predictions for LHC physics. The essential feature
is that the scalar masses and the $A$, $B\equiv B\mu/\mu$ and $\mu$
terms are all generated at the scale of the gravitino $m_{3/2}$. 

However this scenario appears to have cosmological problems (for a
recent discussion of the cosmological gravitino and moduli problem
with references to the earlier literature see \citep{Acharya:2008bk}).
In mSUGRA the gravitino mass is taken to be at the weak scale whereas
to avoid conflicts with the standard Big Bang Nucleosynthesis (BBN)
scenario the gravitino mass (as well as the lightest modulus) should
be heavier than about $10TeV$. This would also entail a cosmological
modulus problem since the scalar partner of the Goldstino (the sGoldstino)
generically has a mass which is of the same order as the gravitino
(see for example \citep{Covi:2008ea}).

\subsection{Sequestered mSUGRA Models}

This class of models is characterized by the cancellation of the leading
terms that contribute to the soft terms. This means that at the high
(GUT?) scale we should have\begin{eqnarray}
m_{\alpha\bar{\beta}}^{2} & = & (m_{3/2}^{2}\tilde{K}_{\alpha\bar{\beta}}-F^{A}F^{\bar{B}}R_{A\bar{B}\alpha\bar{\beta}})\ll m_{3/2}^{2}\tilde{K}_{\alpha\bar{\beta}},\label{eq:mseq}\\
A_{\alpha\beta\gamma}|_{classical} & = & F^{A}D_{A}Y_{\alpha\beta\gamma}\ll m_{3/2}Y_{\alpha\beta\gamma}.\label{eq:Aseq}\end{eqnarray}

The point is that potentially FCNC violating terms generated at the
high (GUT?) scale are suppressed along with the flavor diagonal terms.
The gaugino masses are generated by Weyl anomaly effects even if the
classical terms are zero. The soft scalar masses and the $A$-terms
are then generated through Renormalization Group (RG) running down
to the MSSM scale. Obviously the requirement imposes a restriction
on the moduli dependence of the Kaehler metric of the visible sector
$\tilde{K}_{\alpha\bar{\beta}}$. The question is whether the class
of models where this restriction holds is more natural (or more plausible)
than the mSUGRA restriction discussed in the previous subsection.
What we will argue later is that from the string theory point of view
at least, there is a viable class of models in which this scenario
can be quite explicitly realized.

In this class of models the gaugino masses, the scalar masses, as
well as the $A$ and $B$ terms at the MSSM scale, are of order $(\alpha/4\pi)m_{3/2}\sim10^{-2}m_{3/2}$
for a typical gauge coupling. To get weak scale soft terms then we
need $m_{3/2}\sim10-100TeV$. This will eliminate the cosmological
gravitino problem which afflicts mSUGRA. However in typical realizations
of this scenario there is a potential $\mu$ problem. As can be seen
from \eqref{eq:mu} generically a gravitino mass of $10TeV$ or more
will generate a $\mu$ term which is far too large. However in situations
where sequestering is realized, the leading contributions to the last
two terms of \eqref{eq:mu} will cancel. The question then is whether
the subleading terms will generate a large enough $\mu$ term, when
the $\tilde{\mu}$ term coming from the superpotential \eqref{eq:W}
of the fundamental theory is zero (as is the case in IIB models with
the MSSM on a stack of D3 branes).

\subsection{Gauge Mediated SUSY Breaking (GMSB)\label{sub:Gauge-Mediated-SUSY}}

GMSB is usually discussed within the context of global supersymmetry
- with the gravitino mass and the tuning of the CC tacked on as an
afterthought. But a complete supersymmetric effective theory, valid
at say the GUT scale and below, must necessarily be a SUGRA. The main
argument in favor of GMSB is that, since the hidden sector SUSY breaking
is transmitted by gauge interactions to the visible sector, the soft
terms do not generate FCNC effects. It is obviously crucial then to
suppress potentially flavor violating effects generated by SUGRA effects.
In GMSB models this is effected simply by suppressing the gravitino
mass well below the weak scale - typically it needs to be at the KeV
scale in order to avoid cosmological problems. However this means
that the mechanism of SUSY breaking and transmission becomes more
complicated. Essentially the problem is that the typical modulus in
a SUGRA has Planck scale vacuum expectation values (vev's). However
in GMSB the chiral scalar or scalars in the effective O'Rafferteagh
type models that are responsible for the SUSY breaking, are required
to get vev's which are several orders of magnitude smaller than the
Planck scale. Let us review briefly why this is the case.

Let $X$ be the chiral scalar superfield that is responsible for breaking
SUSY, i.e. at the minimum of the potential $F^{X}|_{0}\ne0$. We can
without loss of generality for the purposes of this discussion, take
this to be a single superfield so that%
\footnote{In the following the subscript $|_{0}$ indicates that the corresponding
quantity is to be evaluated at the relevant local minimum of the potential.%
} $F^{A}|_{0}=0$ for $A\ne X$. The requirement that the CC is zero
means \begin{equation}
V_{0}=|F^{X}|_{0}^{2}-3m_{3/2}^{2}=0,\label{eq:CC=0}\end{equation}
so that $|F^{X}|_{0}=\sqrt{3}m_{3/2}$.%
\footnote{If there are other sources of SUSY breaking which do not couple to
the messengers this is an upper bound.%
} However as discussed above, the way GMSB suppresses possible FCNC
effects coming from SUGRA is by taking $m_{3/2}=e^{K/2}|W|_{0}$ to
be extremely small. i.e. effectively by choosing parameters in the
superpotential such that in Planck units (for instance if the Kaehler
potential is $O(1)$) $|W|_{0}\lesssim10^{-24}$. This is certainly
possible in string theoretic constructions (for example in type IIB
models) and we will come back to this issue later. But this means
that (as we can see immediately from equations \eqref{eq:Mgaugino}\eqref{eq:msoft})
the classically generated soft terms are negligible, and we need some
mechanism for enhancing the relevant connection and curvature components
of the matter metric. Essentially we need a singularity at the origin
of moduli space to enhance the tiny value of $F^{X}$. In typical
GMSB models this is achieved by coupling $X$ to a so-called messenger
sector (with superfields $f,\tilde{f}$ say) taken to be in a vector-like
representation of the standard model gauge group. Thus a term \begin{equation}
\Delta W=Xf\tilde{f},\label{eq:DeltaW}\end{equation}
is added to the superpotential. Below the messenger scale $X_{0}$
one may integrate out the messengers. This gives a threshold effect
at the messenger scale and contributes a term \citep{Giudice:1997ni}
\begin{equation}
\Delta H=\sum_{r=f,\tilde{f}}\frac{T_{i}(r)}{4\pi^{2}}\ln X,\label{eq:DeltaH}\end{equation}
to the gauge coupling function at scales below the messenger scale.
Consequently there is a contribution to the gaugino mass: \begin{equation}
M_{i}=\sum_{r=f,\tilde{f}}T_{i}(r)\frac{g_{i}^{2}}{8\pi^{2}}\frac{F^{X}}{X}.\label{eq:Mgmsb}\end{equation}
Now since $F^{X}\sim m_{3/2}\ll100GeV$ in order to get an acceptable
gaugino mass, the vev of $X$ needs to be suppressed well below its
natural scale in SUGRA, i.e. the Planck scale. A similar enhancement
happens in the curvature term contributing to the soft scalar mass
\eqref{eq:msoft} leading to a flavor diagonal contribution\begin{equation}
m_{\alpha\bar{\beta}}^{2}=m_{0}^{2}K_{\alpha\bar{\beta}},\, m_{0}^{2}=2\sum_{i}c_{i}\left(\frac{\alpha_{i}}{4\pi}\right)^{2}\sum_{r=f,\tilde{f}}T_{i}(r)|\frac{F^{X}}{X}|^{2}.\label{eq:mgmsb}\end{equation}
In various versions of GMSB (direct, indirect, semi-direct, general)
the hidden sector and the messenger sector may undergo modifications/generalizations
so that the effective source of SUSY breaking $F^{X}/X$ may be replaced
by a sum of such terms. But the basic requirement (given that the
$F$ terms are at most of $O(m_{3/2})$) is that the vev's of the
supersymmetry breaking hidden sector field or fields, need to be stabilized
at some scale that is well below the Planck scale. For simplicity
we will continue to assume that there is just the one SUSY breaking
modulus $X$.

Most works on GMSB do not discuss the embedding of the theory within
SUGRA. However as we've argued above, a theory of SUSY breaking cannot
ignore SUGRA. The only attempt at a GMSB discussion within SUGRA that
the author is aware of is that of \citep{Kitano:2006wz}. The model
is defined by the Kaehler potential, \begin{equation}
K=X\bar{X}-\frac{(X\bar{X})^{2}}{\Lambda^{2}}+f\bar{f}+\tilde{f}\bar{\tilde{f}}+K_{MSSM},\label{eq:Kkit}\end{equation}
and superpotential\begin{equation}
W=c+\mu^{2}X+\lambda Xf\tilde{f}+W_{MSSM}.\label{eq:Wkit}\end{equation}
Here the superfields $f,\tilde{f}$ are to be identified as the messengers
of GMSB. The model has a true minimum (i.e. with no tachyons or flat
directions) with the fields taking values $X_{0}=\frac{\sqrt{3}\Lambda^{2}}{6},\, f=\tilde{f}=0$
, provided that $\Lambda^{4}>\frac{12\mu^{2}}{\lambda}$ and the CC
is tuned to zero; i.e.\begin{equation}
\mu^{2}\simeq\sqrt{3}c=\sqrt{3}m_{3/2}.\label{eq:cckit}\end{equation}
Using the standard mass formula for scalar masses in SUGRA the mass
matrices may be evaluated. The scalar messengers have squared masses
\begin{equation}
\frac{\lambda^{2}\Lambda^{4}}{12}\pm\lambda\mu^{2}\label{eq:messmass}\end{equation}
 while the scalar partner of the Goldstino (sGoldstino) has a mass
$m_{X}\simeq2\mu^{2}/\Lambda$. $ $ Finally the SUSY breaking is
characterized by \begin{equation}
F^{X}\simeq\mu^{2}=\sqrt{3}m_{3/2},\label{eq:FX}\end{equation}
so that the relevant mass parameter determining soft terms in GMSB
is (restoring $M_{P}$ for clarity) \begin{equation}
m\sim\frac{\alpha}{4\pi}\frac{F^{X}}{X}\simeq\frac{\alpha}{4\pi}\frac{M_{P}^{2}}{\Lambda^{2}}6m_{3/2}.\label{eq:softGMSB}\end{equation}
This simple model illustrates several features that must generically
be present in GMSB. Firstly, as we discussed before, the gravitino
mass needs to be well below the weak scale in order to suppress the
naturally occurring gravity (moduli) mediated contribution. The first
factor in \eqref{eq:softGMSB}gives a suppression of $O(10^{-2})$,
so if we choose $m_{3/2}\lesssim1KeV$ in order to avoid gravitino
cosmological problems (as is usually done), then we must have a cutoff
$\Lambda\lesssim10^{-5}M_{P}$. Also we need to impose the CC fine
tuning condition \eqref{eq:cckit}.

This scenario has a cosmological modulus problem. One might expect
this \citep{Kane:2010gk} from the expression for the sGoldstino mass
given in Covi et al \citep{Covi:2008ea}, however that assumes that
the only relevant scale is the the Planck scale. With a Kaehler potential
as in \eqref{eq:Kkit} however there is a scale $\Lambda$ which is
significantly lower than the Planck scale. This has the potential
of raising the sGoldstino mass above this bound. Nevertheless as we
will argue below it cannot be raised high enough to evade cosmological
problems.

Using \eqref{eq:FX} we can rewrite the mass of the sGoldstino (see
below eqn.\eqref{eq:messmass}) as \begin{equation}
m_{X}\simeq2\sqrt{3}m_{3/2}\frac{M_{P}}{\Lambda}.\label{eq:mX}\end{equation}
 On the other hand from \eqref{eq:softGMSB}, we have $\Lambda/M_{P}=(6\alpha m_{3/2}/4\pi m)^{1/2}$,
so that from \eqref{eq:mX} we get\begin{equation}
m_{X}=2\sqrt{3}m_{3/2}\left(\frac{4\pi m}{6\alpha m_{3/2}}\right)^{1/2}\sim10\sqrt{m_{3/2}m}.\label{eq:mX2}\end{equation}
In the last relation we have used $\alpha/4\pi\sim10^{-2}$. Even
for the largest allowed gravitino mass $\sim1KeV$, if we take the
soft mass scale $m$ to be at the weak scale ($M_{W}\sim100GeV$),
this gives a modulus mass around $0.1GeV$ which is far too small
to evade the modulus problem%
\footnote{For a recent discussion see \citep{Acharya:2009zt}. %
}. In fact to satisfy the bound on the modulus $m_{X}\gtrsim10TeV$
we would need to take the soft mass scale $m\sim10^{12}GeV$ which
of course would be incompatible with a SUSY solution to the hierarchy
problem.

\subsection{GMSB - mSUGRA comparison}

How does the previous scenario compare with the corresponding mSUGRA
one. Firstly ignoring any fundamental (say string theory based) derivation,
one could take the same starting point as the model \eqref{eq:Kkit}\eqref{eq:Wkit}
except for two ingredients: one does not need the messenger sector,
and the gravitino mass should have a weak scale value i.e. around
$100-1000GeV$. In the next section we will look at string theory
scenarios but here let us focus on using the same SUGRA embedding
as in the GMSB case discussed above. So we take \begin{eqnarray}
K & = & X\bar{X}-\frac{(X\bar{X})^{2}}{\Lambda^{2}}+g(X,\bar{X})k_{\alpha\bar{\beta}}C^{\alpha}\bar{C}^{\bar{\beta}}+[Z_{\alpha\beta}(X,\bar{X})C^{\alpha}C^{\beta}+h.c.]\label{eq:KmSUGRA}\\
W & = & c+\mu^{2}X+\frac{1}{6}\tilde{Y}_{\alpha\beta\gamma}C^{\alpha}C^{\beta}C^{\gamma}+\ldots.\label{eq:Wmsugra}\end{eqnarray}
In the above $\tilde{K}_{\alpha\bar{\beta}},Z_{\alpha\beta}$ , may
of course depend on other moduli (which don't break SUSY) which are
not explicitly written down. Also in $W$ we take the standard model
Yukawa coupling to be independent of $X$ and the $\tilde{\mu}$ term
to be zero. As in the previous discussion the potential will have
a minimum at $X=X_{0}=\frac{\sqrt{3}\Lambda^{2}}{6},\, C=0$, and
the SUSY breaking will be characterized by $F^{X}\simeq\mu^{2}=\sqrt{3}m_{3/2}$
after fine-tuning the CC as before. This scenario is of course a particular
case of the situation we discussed before, and will yield FCNC conserving
scalar masses and trilinear couplings (see equations \eqref{eq:m0mSUGRA}\eqref{eq:A0mSUGRA}).
For example in the simplest case $g=X\bar{X}$ $m_{0}^{2}=m_{3/2}^{2}(1-3\partial_{X}\partial_{\bar{X}}\ln g(X,\bar{X}))=m_{3/2}^{2}$.
However one might expect that the particular structure of the Kaehler
potential (i.e. the form of the third term in \eqref{eq:KmSUGRA})
cannot be preserved when loop corrections are added. The dangerous
terms are the quadratically divergent supergravity loop corrections.
But to one loop order they have been calculated using the Coleman-Weinberg
potential (see for example \citep{Choi:1997de}). With the cutoff
$\Lambda$ the corrections to the squared scalar mass is of order\begin{equation}
\frac{N\Lambda^{2}}{(4\pi)^{2}}m_{3/2}^{2}\sim10^{-4}m_{0}^{2},\label{eq:quantcorr}\end{equation}
where in the last relation we used $N$ the number of chiral scalars
in the loop to be around $10^{2}$ and the cutoff to be $\Lambda\sim M_{GUT}\sim10^{-2}$.
This estimate shows that the FCNC effects generated by these quantum
corrections can be safely ignored since $\Delta m_{FCNC}^{2}/m_{0}^{2}<10^{-3}$.
Similarly any FCNC effect in the trilinear couplings coming from quantum
corrections is suppressed. Thus the boundary values used in mSUGRA
are safe from large quantum corrections at the high scale, and as
is well known the logarithmic RG evolution down to the weak scale
will not generate any large FCNC effects. 

The upshot is that once the embedding into supergravity is considered,
there is no particular advantage in choosing GMSB over mSUGRA. Both
require an ansatz about the coupling of MSSM visible sector to the
supersymmetry breaking hidden sector. In GMSB one postulates an additional
sector (the so-called messenger sector) which couples directly to
the SUSY breaking sector and communicates the SUSY breaking via gauge
interactions, which are of course naturally flavor diagonal. However
SUGRA effects, which are always present, need to be suppressed way
below the weak scale by tuning the gravitino mass to be extremely
small. On the other hand in mSUGRA to get FCNC conserving initial
conditions for the soft parameters, one needs to assume a particular
type of coupling of the SUSY breaking moduli to the MSSM Kaehler metric.
This may indeed be affected by quantum corrections at the high scale.
However if the effective cutoff is well below the Planck scale (as
is required to be the case in GMSB too) then these corrections are
negligible. Furthermore one needs to tune the gravitino mass only
to the weak scale, and this immediately implies a natural solution
to the $\mu$ and $B\mu$ problem, a feature which is absent in GMSB.

\section{String theory considerations}

String theory is supposed to be the ultra-violet completion of supergravity.
Of course not every SUGRA may have such a completion so it is natural
to favor those effective supergravity theories that have such a completion.
Let us briefly review the general structure of the SUGRA that would
emerge from a string theory compactified to 4 dimensions.

In both heterotic and type II models (compactified on a Calabi-Yau
manifold $Y$) the Kaehler potential takes the the form

\begin{equation}
K=-2\ln\left({\cal V}\right)-\ln\left(i\int\Omega\wedge\bar{\Omega}\right)-\ln(S+\bar{S}).\label{eq:Kstring}\end{equation}
Here ${\cal V}$ is the volume of $Y$and depends on the Kaehler moduli,
$\Omega$ is the holomorphic three-form on $Y$ which depends on the
complex structure moduli, and $S$ is the dilaton-axion superfield
whose real part essentially defines the string coupling. 

As pointed out in \citep{Kaplunovsky:1993rd} there are two typical
cases: a) $F^{S}\gg F^{M}$, or b)$F^{M}\gg F^{S}$, where $M$ is
a modulus. In heterotic compactifications the axionic shift symmetry
associated with $S$ ensures that the superpotential is independent
of it (except non-perturbatively). In this case one naturally gets
mSUGRA soft terms with $m_{0}\sim A_{0}\sim M_{i}\sim m_{3/2}$ thus
giving mSUGRA if case a) is realized.%
\footnote{Estimates of string loop corrections to this dilaton dominated scenario
are given in \citep{Louis:1994ht}. These imply that when FCNC constraints
are taken into acount the gravitino mass should be raised to about
400GeV. This implies a small hierarchy problem, but this is in any
case there in all SUSY mediation mechanisms because of the experimental
lower bounds on the chargino and Higgs masses. %
} Unfortunately there is no known moduli stabilization mechanism that
achieves case a).

In actual realizations of string theoretic SUGRA the opposite situation
is what is obtained - i.e. in all known string compactifications (with
fluxes and non-perturbative terms) which stabilize all the moduli,
one finds that $F^{M}\gg F^{S}$ for at least one modulus. This is
true of SUSY breaking in both heterotic and type IIB models that have
been studied so far, and it is possibly a generic feature of string
theoretic SUGRA. Thus it seems that the simple dilaton dominated SUSY
breaking scenario, and hence the hope of having a model independent
justification for mSUGRA discussed in the previous paragraph, is hard
to obtain in string theory. Faced with this situation there are two
options that one can pursue. 

\begin{enumerate}
\item Choose flux configurations such that the gravitino mass is well below
the weak scale and use GMSB. 
\item Use large volume compactifications as in \citep{Balasubramanian:2005zx}
(LVS).
\end{enumerate}
There are several issues that need to be addressed before one can
claim to have a viable string theory realization of GMSB. First one
needs to tune the fluxes so that the gravitino mass is well below
the weak scale. In typical GMSB models this is effectively a tuning
of the SUSY breaking scale to be a factor of at least $10^{6}$ below
the SUSY breaking scale of mSUGRA or LVS models. In the landscape
of string theory (since the frequency of models with zero CC and broken
SUSY goes as $F^{6}$ \citep{Douglas:2004qg}) this is less likely
by factor of $10^{36}$! Even after selecting such a class of models
one needs a supersymmetry breaking chiral scalar field $X$ which
must have a vev that is much smaller than the Planck scale (see above
discussion after (\ref{eq:softGMSB})). However in the (LVS) string
theory regime in which meaningful calculations can be done with current
technology (i.e. where the KK scale is well below the string scale
which in turn is below the Planck scale), all moduli as well as the
dilaton have vev's which are larger than the Planck scale. Thus we
need another sector, the one represented by $X$ in subsection \eqref{sub:Gauge-Mediated-SUSY}.

In a string theory embedding one might think of supersymmetrically
integrating out all the string theory moduli at a scale that is higher
than the messenger scale $M_{messenger}\sim X_{0}\sim\Lambda^{2}/M_{P}$,
so that below this scale the system is well described by a model such
as the one discussed in \citep{Kitano:2006wz} (see discussion in
subsection \ref{sub:Gauge-Mediated-SUSY}). Preliminary investigations
\citep{deAlwis:2010tp} however seem to indicate that in order to
produce a scale $\Lambda$ which is parametrically smaller than the
Planck/String scale $\Lambda/M_{P}\lesssim10^{-5}$ as is required
in GMSB, we need extremely large rank ($N>10^{4}$) gauge groups to
produce the requisite non-perturbative contribution to the superpotential
that stabilizes the volume modulus at a large enough value. On the
other hand the LVS scenario \citep{Balasubramanian:2005zx} does produce
an exponentially large volume. However it also breaks SUSY dominantly
in the volume modulus direction, and the corresponding light modulus
(i.e. the sGoldstino) is lighter than the gravitino, so that it cannot
be integrated out in GMSB where the gravitino is the LSP. Thus it
appears that one has to discuss the SUGRA potential involving at least
the lightest modulus (typically the volume modulus) and the field
$X$, and ensure that there is a minimum of the potential in the large
moduli/dilaton region, which however yields a small value for the
field $X$, even though they get F-terms of the same order i.e. $F\sim m_{3/2}M_{P}$.
These requirements cannot be satisfied (see for example \citep{deAlwis:2007qx})
without fine-tuning. 

The alternative is the LVS scenario of SUSY breaking \citep{Balasubramanian:2005zx}\citep{Conlon:2008wa}\citep{Blumenhagen:2009gk}\citep{deAlwis:2009fn}.
Here the classical soft masses (and hence also the FCNC effects) are
highly suppressed by powers of the large compactification volume,
relative to the gravitino mass. All the moduli are stabilized by a
combination of fluxes and non-perturbative effects. The gaugino masses
are then generated by Weyl anomaly (AMSB) effects while the leading
contribution to the soft scalar masses and the $A$ term are generated
by RG running effects \citep{deAlwis:2009fn}. With a soft mass scale
at or below a TeV we need a volume which is at least $10^{5}$ times
the Planck volume in order that FCNC effects are sufficiently suppressed.
The quantum effects are also suppressed relative to the classical
terms by arguments similar to those given earlier (see eqn. (\ref{eq:quantcorr})
and references \citep{deAlwis:2009fn,deAlwis:2008kt},\citep{Burgess:2010sy}).

A brief comment on the phenomenology of M-theory compactifications
on G2 manifolds (see \citep{Acharya:2008zi} and references therein)
is in order here. The soft terms are proportional to $m_{3/2}$ as
in mSUGRA, however the gravitino mass is taken to be greater than
$10TeV$ in order to avoid cosmological problems. Of course now the
little hierarchy problem is somewhat worse (with a fine-tuning of
at least one part in $10^{4}$) and there is still an FCNC problem
unless one makes a special ansatz as in mSUGRA. 

Actually if one is willing to worsen the little hierarchy problem
somewhat, one could solve the cosmological modulus problem within
the LVS string theory derived inoAMSB scheme. According to \citep{Acharya:2008bk}
the upper bound on the gravitino mass is $m_{3/2}\sim500TeV$. If
we take this value then the resulting soft masses in inoAMSB are $(\alpha/4\pi)m_{3/2}\sim2-3TeV$.
This obviously increases the little hierarchy fine-tuning to one part
in $10^{3}$, a factor of about 25 worse than in inoAMSB with $100TeV$
gravitino mass, but somewhat better than what one would have in the
$G_{2}$ case above! The advantage of this $m_{3/2}=500TeV$ version
of inoAMSB is that now (as argued in \citep{deAlwis:2009fn}) the
lower bound on the volume coming from the need to suppress FCNC is
(in Planck units) ${\cal V}\gtrsim10^{4}$ \citep{deAlwis:2009fn}.
This means that the string scale $M_{string}\sim M_{P}/\sqrt{{\cal V}}$
can be as large as $10^{16}GeV$ allowing gauge unification. Most
importantly if we use this smallest allowed volume, the light modulus
mass (i.e. the mass of the scalar partner of the Goldstino) is $M_{sg}\sim m_{3/2}/\sqrt{{\cal V}}\sim5TeV$
thus posssibly evading the cosmological modulus problem.

\section{Conclusions}

We have argued that any viable theory of SUSY breaking and mediation
must be addressed within the SUGRA context and that conclusions drawn
from a purely global analysis may not hold once the full implications
of SUGRA are considered. The most important aspect that is missing
from a purely global analysis is the issue of stabilizing the chiral
scalar fields that are responsible for the supersymmetry breaking
in such a way that the CC is (almost) zero. When these considerations
are taken into account we argued that there is no reason to prefer
GMSB over gravity or moduli mediated SUSY breaking in its various
forms. If one also demands that such a supergravity be embedded in
string theory, it seems that, while it is difficult to realize GMSB,
and it is not clear whether the mSUGRA scenario can be realized either,
the LVS type compactifications with anomaly and gaugino mediation,
which give a viable phenomenology, can be obtained. The following
table gives a rough comparison of these three main mechanisms highlighting
the problems of each of them.

Table: Comparison Chart

\begin{tabular}{|c|c|c|c|}
\hline 
\noalign{\vskip\doublerulesep}
 & GMSB & mSUGRA & Sequestered\tabularnewline
\hline
\hline 
gravitino mass & $m_{3/2}\ll M_{W}$ & $m_{3/2}\gtrsim M_{W}$ & $m_{3/2}\gg M_{W}$\tabularnewline
\hline 
Mediation & gauge & gravitational & anomaly/gaugino\tabularnewline
\hline 
FCNC & natural  & needs special ansatz & natural\tabularnewline
\hline 
$\mu/B\mu$ & problematic & natural & possibly problematic\tabularnewline
\hline 
cosmo gravitino  & OK & problematic & OK\tabularnewline
\hline 
cosmo modulus & problematic & problematic & problematic?\tabularnewline
\hline
\hline 
\noalign{\vskip\doublerulesep}
String embedding & hard & possible? & LVS example\tabularnewline
\hline
\noalign{\vskip\doublerulesep}
\end{tabular}

Several comments are in order. Firstly FCNC suppression in GMSB is
natural in that the main mechanism for mediating SUSY breaking are
gauge interactions - however this requires the suppression of the
SUGRA mediation effects, and this is effected by looking at theories
which have an extremely small gravitino mass. In sequestered theories
the dominant mediation mechanism is the Weyl anomaly and gaugino mediation
and the suppression of direct SUGRA effects is achieved in (LVS) compactifications
of type IIB string theory by the large volume suppression of direct
gravity coupling effects. The $\mu$ problem has been designated as
possibly problematic since the $\mu$-term depends on the terms which
are responsible for the uplift of the CC, and though it is plausible
that they may be generated at the right order, there is no precise
calculation demonstrating that. The lightest modulus may also be problematic
in this scenario unless one increases the gravitino mass to about
$500TeV$, but this would worsen the little hierarchy problem .

Ultimately the correct mechanism may have to be decided by experiment.
Nevertheless it is worthwhile investigating the theoretical questions
posed in the last row of the table above. In particular while currently
the only framework which allows a string theoretic description is
the sequestered case in the last column, a definitive statement about
the first two from the point of view of string theory, would give
us valuable insights into the nature of physics close to the Planck
scale, if the LHC reveals to us the existence of low scale SUSY and
the structure of the soft SUSY breaking terms.

\section{Acknowledgements}

I wish to thank the organizers of the KITP work shop ``Strings at
the LHC and in Early Cosmology'' where part of this work was done.
Some of these ideas were presented at the discussion session on ``Comparing
SUSY breaking and Mediation Mechanisms'' and I thank the participants
for a stimulating discussion. I also wish to thank Howie Baer and
Gordy Kane for discussions on the cosmological modulus problem and
the former for comments on the manuscript as well. This research was
supported in part by the National Science Foundation under Grant No.
NSF PHY05-51164 and by the United States Department of Energy under
grant DE-FG02-91-ER-40672.

\bibliographystyle{apsrev}
\bibliography{myrefs}

\end{document}